\documentclass[12pt,preprint]{aastex}
\begin{document}

\title{GRB~140423A: A Case of Stellar Wind to Interstellar Medium Transition in the Afterglow}

\author{Long Li\altaffilmark{1,2}, Xiang-Gao Wang\altaffilmark{1,2}, WeiKang Zheng\altaffilmark{3}, Alexei S. Pozanenko\altaffilmark{4,5,6}, Alexei V. Filippenko\altaffilmark{3,7}, Songmei Qin\altaffilmark{8}, Shan-Qin Wang\altaffilmark{1,2}, Lu-Yao Jiang\altaffilmark{1,2}, Jing Li\altaffilmark{1,2}, Da-Bin Lin\altaffilmark{1,2}, En-Wei Liang\altaffilmark{1,2}, Alina A. Volnova\altaffilmark{4}, Leonid Elenin\altaffilmark{9}, Evgeny Klunko\altaffilmark{10}, Raguli Ya. Inasaridze\altaffilmark{11}, Anatoly Kusakin\altaffilmark{12}, and Rui-Jing Lu\altaffilmark{1,2}}

\altaffiltext{1}{Guangxi Key Laboratory for Relativistic Astrophysics, School of Physical Science and Technology, Guangxi University, Nanning 530004, China; wangxg@gxu.edu.cn}
\altaffiltext{2}{GXU-NAOC Center for Astrophysics and Space Sciences, Nanning 530004, China}
\altaffiltext{3}{Department of Astronomy, University of California, Berkeley, CA 94720-3411, USA; weikang@berkeley.edu, afilippenko@berkeley.edu}
\altaffiltext{4}{Space Research Institute of RAS, Profsoyuznaya, 84/32, Moscow 117997, Russia}
\altaffiltext{5}{National Research University Higher School of Economics, Myasnitskaya 20, 101000, Moscow, Russia}
\altaffiltext{6}{Moscow Institute of Physics and Technology (MIPT), Institutskiy Pereulok, 9, Dolgoprudny, 141701, Russia}
\altaffiltext{7}{Miller Senior Fellow, Miller Institute for Basic Research in Science, University of California, Berkeley, CA 94720, USA}
\altaffiltext{8}{Mathematics and Physics Section, Guangxi University of Chinese Medicine, Nanning 53001,China}
\altaffiltext{9}{Keldysh Institute of Applied Mathematics, Miusskaya sq., 4, Moscow, 125047, Russia}
\altaffiltext{10}{Institute of Solar Terrestrial Physics, Irkutsk, 664033 Russia}
\altaffiltext{11}{Kharadze Abastumani Astrophysical Observatory, Ilia State University, Tbilisi, 0162, Georgia}
\altaffiltext{12}{Fesenkov Astrophysical Institute, Almaty, 050020, Kazakhstan}

\begin{abstract}

We present very early ground-based optical follow-up observations of GRB~140423A, which was discovered by \emph{Swift}/BAT and by {\it Fermi}/GBM. Its broadband afterglow was monitored by {\it Swift}/XRT and ground-based optical telescopes from $T_0+$70.96~s to 4.8~d after the {\it Swift}/BAT trigger. This is one more case of prompt optical emission observation. The temporal and spectral joint fit of the multiwavelength light curves of GRB 140423A reveals that achromatic behavior is consistent with the external shock model including a transition from a stellar wind to the interstellar medium (ISM) and energy injection. In terms of the optical light curves, there is an onset bump in the early afterglow with a rising index $\alpha_{\rm O,I} = -0.59 \pm 0.04$ (peaking at $t_{\rm peak}-T_0 \approx 206$~s). It then decays with a steep index $\alpha_{\rm O,II} = 1.78 \pm 0.03$, and shows a steeper to flatter ``transition" with $\alpha_{\rm O,III} = 1.13 \pm 0.03$ at around $T_0 + 5000$~s. The observed X-ray afterglow reflects an achromatic behavior, as does the optical light curve. There is no obvious evolution of the spectral energy distribution between the X-ray and optical afterglow, with an average value of the photon index $\Gamma \approx 1.95$. This ``transition" is consistent with an external shock model having the circumburst medium transition from a wind to the ISM, by introducing a long-lasting energy injection with a Lorentz factor stratification of the ejecta. The best parameters from Monte Carlo Markov Chain fitting are $E_{\rm K,iso} \approx 2.14\times10^{55}$ erg, $\Gamma_0 \approx 162$, $\epsilon_e \approx 0.02$, $\epsilon_B \approx 1.7\times10^{-6}$, $A_\ast \approx 1.0$, $R_t \approx 4.1\times10^{17}$ cm, $n \approx 11.0 \rm\ cm^{-3}$, $L_0 \approx 3.1\times10^{52} \rm\ erg\ s^{-1}$, $k \approx 1.98$, $s \approx 1.54$, and $\theta_j > 0.3$ rad.
\end{abstract}


\section{Introduction}

Gamma-ray bursts (GRBs) are extremely energetic explosions in the Universe. When the initial burst of gamma rays subsides, a longer-lived ``afterglow" is normally emitted at longer wavelengths (X-ray, ultraviolet, optical, infrared, microwave, and radio) \citep[e.g.,][]{zhang04,Kumarzhang15,Warren17,Zhang18}. This afterglow can be well explained using the synchrotron emission caused by collisions between the ultrarelativistic jet and the circumburst medium \citep{meszaros97}, while some extended or flat behavior in X-rays could be caused by late activity of the central engine \citep{Komissarov09,Barkov10}. The temporal and spectral evolution of the multiwavelength afterglow can be used to diagnose the underlying radiation mechanism and the profile of the circumburst medium \citep{sari98}. Most observed events have a duration $> 2$~s and are classified as long gamma-ray bursts.

It is generally believed that almost every long GRB is associated with the death of a massive star \citep{woosley93,MacFadyen99,woosley06}, especially those with stripped envelopes
\citep[e.g.,][]{MacFadyen99}. The circumburst medium has likely been impacted by the stellar wind created by the progenitor, and it can be described as $n(r) \propto r^{-k}$. Some authors investigate the case of a wind ($k = 2$; \citealt{dai98,vink00,dai03,chevalier04,vink05,xin12}), while other authors (e.g., \citealt{panaitescu02}) favor a homogeneous density. However, \citet{liang13} and \citet{yi13} take GRB onset bumps as probes of the properties of the ambient medium, finding that the density profile can be described as $n \propto r^{-1}$, in which the index ($k = 1$) is between that of a homogeneous medium ($k = 0$) and a stellar wind ($k = 2$).

Since the progenitors and the circumburst medium are embedded in the interstellar medium (ISM), there should eventually be a transition between the circumburst medium and the ISM. The interaction between a stellar wind and the surrounding ISM could create a bubble structure \citep{weaver77}. In this scenario, the stellar wind terminates at a radius $R_t \approx 10^{18}$--$10^{20}$ cm, where the density jumps by a factor of 4 or more, with the lower value expected for an adiabatic shock \citep{peer06}. \cite{dai03} explain the early afterglow of GRB~030226 by considering an ultrarelativistic collimated blast wave expanding in a density-jump medium that forms a reverse shock and a forward shock; they conclude that the density-jump formation is the result of the interaction of a stellar wind from a massive star and its outer environment. \cite{kamble07} argue that the early X-ray afterglow of GRB~050319 suggests a wind-like density profile of the circumburst medium while the late-time optical afterglow was in accord with evolution in the ISM; different microphysics parameters in the wind-like medium and the ISM medium were considered.

\cite{jin09} suggested that the afterglow of GRB 081109A is generated in a wind bubble structure, with the optical and X-ray afterglows falling in different radiation regions separated by characteristic frequencies. \cite{kong10} state that the rebrightenings in the afterglows of GRB 060206, GRB 070311, and GRB 071010A are caused by variations of the microphysics in the wind bubbles. \cite{fraija17} suggest that the early-time light curve of afterglow of GRB 160625B might be powered by the interaction between a jet and the wind-like medium while the late-time afterglow can be powered by the jet and the ISM, indicating that there is a transition from the stellar wind to the ISM. \cite{fraija19} show that long-lasting multiwavelength observations of GRB 190114C are consistent with the standard synchrotron forward-shock model that evolves from a stratified stellar-wind-like medium to a uniform ISM-like medium.

In practice, we can use the afterglow light curve to determine the properties of the density profile of the circumburst medium. Here we report our observations of an optical afterglow of GRB~140423A, model the light curve of the X-ray afterglow observed by the X-Ray Telescope (XRT) onboard the \emph{Neil Gehrels Swift Observatory} (\emph{Swift}; \citealt{gehrels04}) as well as the optical afterglow light curves, and explain the peculiar features shown in the optical light curves. Observations and data reduction are reported in Section \ref{sec:obs}, and Section \ref{sec:data} presents an analysis of the optical and X-ray data. The modeling method and results are presented in Section \ref{sec:results}, and we provide a discussion and conclusions in Section \ref{sec:dis}. Throughout, the notation $Q_{n}=Q/10^{n}$ is used for the physical parameters in cgs units, and the temporal and spectral slopes are defined as $F\propto t^{-\alpha} \nu^{-\beta}$. The confidence level of the uncertainties is $1 \sigma$. A concordance cosmological model with H$_0 = 69.6 \rm\ km\ s^{-1}\ Mpc^{-1}$, $\Omega_M = 0.286$, and $\Omega_\Lambda = 0.714$ is adopted.

\section{Observations and Data Reduction}
\label{sec:obs}

GRB 140423A was triggered (trigger 596901) by the Burst Alert Telescope (BAT) onboard \emph{Swift} on April 23, 2014 at 08:31:53 UT (denoted as $T_0$ in this paper) and also (trigger 419934761/140423356) by the Gamma-Ray Burst Monitor (GBM) onboard {\it Fermi} on April 23, 2014 at 08:32:38.54 (UT dates are used throughout this paper). The burst was also detected by Konus-Wind  in the waiting mode \citep{Golenetskii14}. Both the BAT and GBM light curves show multiple-peak structure with $T_{90} = 134$~s and $95$~s, respectively. The X-ray Telescope (XRT) onboard \emph{Swift} began observing the X-ray afterglow of GRB 140423A at 2943.5~s after the BAT trigger (\citealt{Burrows14}; as shown in Figure 1).

We downloaded the \emph{Fermi}/GBM data for GRB 140423A from {\it Fermi} Archive FTP website\footnote{ftp://legacy.gsfc.nasa.gov/fermi/data/}. The Gamma-Ray Spectral Fitting Package (RMFIT, version 4.3.2)\footnote{http://fermi.gsfc.nasa.gov/ssc/data/analysis/rmfit/} was used to extract the time-integrated (from $T_0-66.561$ s to $T_0+28.672$ s) spectra. Two NaI detectors (n6, n9) and one BGO detector (b1) having the smallest angle to the source were selected. As shown in Figure 2, the GBM time-integrated spectrum of GRB~140423A can be fitted well by a Band function \citep{band93} with $E_{\rm p} = 111.0 \pm 16.1$ keV, low-energy photon spectral index $\alpha = -0.14 \pm 0.23$, and high-energy photon spectral index $\beta = -1.90 \pm 0.08$. The adopt $\chi^2$ statistics when fitting, and $\chi^2 = 458.7$ with 363 degrees of freedom, thus giving reduced $\chi^2 = 1.26$. The observed $\gamma$-ray fluence in the energy band 10--1,000 keV is $S_\gamma=(2.15 \pm 0.30) \times 10^{-5}$ erg cm$^{-2}$. The isotropic $\gamma$-ray energy $E_{\rm{\gamma,iso}}$ can be derived from $S_\gamma$ according to
\begin{equation}
E_{\rm{\gamma,iso}}=\frac{4 \pi D_L^2 k S_\gamma}{1+z},
\end{equation}
where $D_{\rm L}$ is the luminosity distance of the source at redshift $z$, and the
parameter $k$ is a factor to correct the observed $\gamma$-ray energy
in a given bandpass to a broad band (e.g., 1-10$^{4}$ keV in the rest
frame) with the observed GRB spectra \citep{Bloom01}. We obtained $E_{\rm \gamma, iso}=(6.54 \pm 0.91) \times 10^{53}$ erg for GRB 140423A at redshift $z=3.26$ \citep{tanvir14}. GRB 140423A fits into the well-known relationship between $ E_p $ and $ E_{\rm \gamma,iso} $ for long duration GRBs \citep{Minaev20}.

Dozens of ground-based optical telescopes observed the optical afterglow of GRB~140423A \citep{2014GCN.16318....1V,2014GCN.16272....1S,2014GCN.16264....1V,2014GCN.16247....1V,
2014GCN.16185....1B,2014GCN.16175....1H,2014GCN.16174....1B,2014GCN.16173....1F,2014GCN.16170....1L,
2014GCN.16169....1C,2014GCN.16165....1H,2014GCN.16160....1K,2014GCN.16153....1C,elenin14,
2014GCN.16145....1F,2014GCN.16176....1O,2014GCN.16171....1G,2014GCN.16167....1T,2014GCN.16166....1D,
2014GCN.16164....1P,2014GCN.16163....1A,zheng14,2014GCN.16154....1X,2014GCN.16151....1M,
2014GCN.16144....1K}. 

The 0.76-m Katzman Automatic Imaging Telescope (KAIT) at Lick Observatory responded automatically to the \emph{Swift} GRB 140423A trigger and began imaging the field on April 23 at 08:33:03, $T_0+$70.96~s later. In the first step of the observations, 10 short-exposure images in the \emph{Clear} filter\footnote{\emph{Clear} filter means unfiltered frames. \cite{Li03} show that unfiltered observations mostly mimic $R$-band photometriy, so we regard all of the unfiltered data in this paper as $R$ when empirically analyzing and model fitting.} with 1~s duration were taken from $T_{0}+ 70.96$ to 100.96 s, which covers the prompt emission phase  and do not display any correlation with prompt emission in gamma-ray domain. Subsequently, the observations were performed with an automatic sequence in the \emph{V}, \emph{I}, and \emph{Clear} filters, and the exposure time was 20~s per image \citep{zheng14}. The 0.4-m telescope ORI-40 of the ISON-NM Observatory also monitored the field starting on April 23 at 09:01:03; unfiltered images of 30~s were taken and a fading optical source was detected \citep{elenin14}. The 1-m Zeiss-1000 (East) telescope of the Tien Shan Astronomical Observatory (TShAO) took images in the $R$ filter of 300~s and 540~s duration on April 23 at 15:04:19 and 19:04:44 (respectively), and on April 24 at 15:20:01 \citep{GCN16318}.  The 0.7-m AS-32 telescope of the Abastumani Observatory (AAO) obtained unfiltered frames with exposures of 120~s on April 24 00:29:00--01:16:00 and 17:46:00--19:16:00 \citep{GCN16264}. The 1.5-m AZT-33IK telescope of the Sayan Observatory (Mondy) observed GRB 140423A in the $R$ filter with a 60~s exposure on April 25 and 120~s exposures on April 26 and 27 \citep{GCN16247}.

The optical data reduction was carried out following standard routines in the Image Reduction and Analysis Facility (IRAF)\footnote{IRAF is distributed by the National Optical Astronomy Observatory (NOAO), which is operated by AURA, Inc., under a cooperative agreement with the NSF; more details can be seen at http://iraf.noao.edu/.} package. All of our optical observations are reported in Table 1 and the afterglow light curves are shown in Figure 3. The \emph{Swift}/XRT light curves were extracted from the UK \emph{Swift} Science Data Center at the University of Leicester \citep{evans09}\footnote{https://www.swift.ac.uk/burst\_analyser/00596901/}, and also shown in Figure 3.

\section{Analysis of the Afterglow of GRB 140423A}
\label{sec:data}

\subsection{Temporal and Spectral Behavior of the Afterglow}
To get the temporal profile of the GRB 140423A afterglow,
we employed a single power-law (SPL) function \citep[e.g.,][]{liang08,Pozanenko13,wang15}

\begin{equation}
F = F_0 t^{ -\alpha},
\end{equation}
\noindent
where $F_0$ is the flux normalization and $\alpha$ is the afterglow flux decay index,
and also a broken power-law function,

\begin{equation} F=F_1\left [
\left (   \frac{t}{t_b}\right)^{\omega\alpha_1}+\left (
\frac{t}{t_b}\right)^{\omega\alpha_2}\right]^{-1/\omega},
\end{equation}
\noindent
where $F_1$ is the flux normalization, $\alpha_{1}$ and $\alpha_{2}$ are respectively the afterglow flux decay indices before and after the break time ($t_b$), and $\omega$ is a smoothness parameter which represents the sharpness of the break. Figure 3 shows the X-ray afterglow, which was observed starting about 3000~s after the BAT trigger; it can be fitted with a SPL function having indices of $\alpha_{\rm X,II}=1.78 \pm 0.03$ and $\alpha_{\rm X,III}=1.14 \pm 0.03$ before and after $\sim 5000$~s, respectively. For the optical afterglow, we can see that clear smooth onset bumps at early epochs rise with an index $\alpha_{\rm O,I} = -0.59 \pm 0.04$ and peak at $t_{\rm peak}-T_0 \approx 206$~s, and then decay with a steep index $\alpha_{\rm O,II} = 1.78 \pm 0.03$. The optical light curves exhibit a steeper to flatter ``transition" with $\alpha_{\rm O,III} = 1.13 \pm 0.03$ at around 5000~s. We do not detect the jet break feature in the optical light curve up to the end of observations on 4.8 days.

The \emph{Swift}/XRT spectra are taken from the UK \emph{Swift} Science Data Center at the University of Leicester \citep{evans09}\footnote{https://www.swift.ac.uk/xrt\_spectra/00596901/}. We also analyze the spectral energy distributions (SEDs) of the GRB 140423A afterglow by jointly fitting the optical and XRT data with the Xspec package \citep{arnaud96}. The XRT data are corrected for the photoelectric absorption of hydrogen in our Galaxy and the host galaxy, and for the extinction of optical data owing to dust grains in our Galaxy and the host galaxy. The equivalent hydrogen column density of our Galaxy is $N_{\rm H} = 1.09 \times 10^{20}$ cm$^{-2}$; that of the host galaxy, $N_{\rm H}^{\rm host} = (4.22 \pm 2.95) \times 10^{21}$ cm$^{-2}$, is derived from the time-integrated XRT spectrum. The extinction in our Galaxy is $A_V = 0.030$ mag, $A_R = 0.019$ mag, and $A_I = 0.016$ mag in the burst direction \citep{schlafly11}. The extinction caused by dust grains in the host galaxy is characterized by the extinction curves of the Small Magellanic Cloud (SMC), whose standard value for $R_V$ (the ratio of the total to selective extinction) is $R_{V,{\rm SMC}} = 2.93$ \citep{pei92}.

The results of the SED fitting are shown in Figure 4. We divided the joint optical and XRT spectra of the afterglow into two time intervals, before and after the ``transition" time: 3000--5000~s and 15,000--40,000~s. The SED of the joint optical and X-ray spectrum can also be well fitted with a SPL function, with photon indices $\Gamma_{\rm I} = 1.96 \pm 0.07$ and $\Gamma_{\rm II} =1.94 \pm 0.08$. We regard $\Gamma = 1.95\pm 0.08$ as an average value of the photon index during the afterglow phase, and spectral indices are $\beta=\Gamma - 1 = 0.95\pm 0.08$. From the two time-resolved XRT spectra, we derived the color excess of the host galaxy to be $E_{\rm I}(B-V) = 0.18 \pm 0.05$ mag and $E_{\rm II}(B-V) = 0.19 \pm 0.06$ mag.

\subsection{The Wind to ISM Transition in GRB 140423A}
\label{sec:analysis}
We test the ``Closure relation'' in various afterglow models (as shown in Table 2; more details are provided by \citealt{zhang06} and \citealt{gao13}) using the observational temporal index $\alpha$ and spectral index $\beta$ of GRB 140423A. The temporal results illustrate that $\bigtriangleup \alpha_{X,O} = \alpha_{\rm X,III}-\alpha_{\rm O,III} \approx 0$, and spectral results display that $\bigtriangleup \beta_{X,O}= \beta_{X}-\beta_{O} = 0$, indicating that the X-ray and optical afterglows are located in the same spectral regimes \citep[][]{wang15}, e.g., $\nu_m < \nu_O < \nu_X < \nu_c$ or $\nu_m < \nu_c < \nu_O < \nu_X$, where $\nu_m$ and $\nu_c$ are (respectively) the minimum injection frequency and cooling frequency for synchrotron radiation, and $\nu_{\rm O}$ and $\nu_{\rm X}$ are (respectively) the optical frequency and the X-ray frequency. Moreover, the optical light curves show a steeper to flatter ``transition" with $\bigtriangleup \alpha_{O} = \alpha_{\rm O,III}-\alpha_{\rm O,II}=-0.65$ at around 5000 s. After testing the Closure relation in various afterglow models, we argue that the ``transition" was caused by the cicumburst medium which transitioned from a density distribution $n(r) \propto r^{-2}$ (stellar wind) to $n =$ constant (homogeneous ISM). To satisfy the Closure relation, energy injection needs to be considered. A slow-cooling regime located in $\nu_m < \nu_{\rm O} < \nu_{\rm X} < \nu_c$ and $\beta = (p-1)/2$ ($p$ is the index of the synchrotron radiating electron spectrum $N_e \propto \gamma_e^{-p}$) with energy injection was selected. Note that the temporal decay index will not change when the circumburst medium transitions from wind to ISM in the spectral regime $\nu_m < \nu_c < \nu_O < \nu_X$.

There are three possible physical origins for the energy injection. The first is the central engine itself lasting longer \citep[e.g.,][]{Dailu98,Zhang&Meszaros01}, and its behavior can be described as a power-law luminosity history $L(t)=L_{0}(t/t_{s})^{-q}$, in which $q$ is the energy-injection parameter and requires $q<1$ \citep[][]{Zhang&Meszaros01,zhang06}. The second is that the central-engine activity may be brief, but at the end of the prompt phase the ejecta have an energy injection with a Lorentz factor stratification \citep{rees98}. The amount of ejected mass moving with Lorentz factors greater than $\Gamma$ is $M(>\gamma) \propto \gamma^{-s}$, and the mass is added to the blastwave when the blastwave progressively decelerates with $E\propto\gamma^{1-s}\propto\Gamma^{1-s}$; $s$ is the energy-injection parameter, $\gamma$ is the Lorentz factor of the ejecta, and $\Gamma$ is the Lorentz factor of the blastwave. Since the energy is injected when $\Gamma \approx \gamma$, the reverse shock is very weak, and one can neglect the reverse shock contribution \citep{Kumarzhang15}. Only when $s>1$ does one expect a change in the fireball dynamics \citep{zhang06,rees98,panaitescu98,sari00}. The third is that the energy injection is also brief, but the outflow has a significant fraction of the Poynting flux \citep[e.g.,][]{Usov92,Thompson&Duncan94,lyutikov03,kobayashi05,zhang06,zhang11,Pozanenko13}. In our case of GRB 140423A, there is no obvious evidence that the outflow has a significant fraction of the Poynting flux. Furthermore, the energy injection lasts from $10^2$~s to $10^6$~s, and thus the second scenario was chosen.

For the first energy injection scenario, the Lorentz factor evolution is $\Gamma\propto R^{-\frac{2+q}{4-2q}}\propto t_{\rm obs}^{-\frac{2+q}{8}}$ (ISM) and $\Gamma\propto R^{\frac{q}{2q-4}}\propto t_{\rm obs}^{-\frac{q}{4}}$ (wind). The Closure relations with $q$ are also listed in Table 2. For the second energy injection scenario, one also can have the Lorentz factor evolution $\Gamma\propto R^{-\frac{3}{1+s}}\propto t_{\rm obs}^{-\frac{3}{7+s}}$ (ISM) and $\Gamma\propto R^{-\frac{1}{1+s}}\propto t_{\rm obs}^{-\frac{1}{3+s}}$ (wind).  The injection mechanisms can be considered equivalent for the first and second scenarios, as far as the blastwave dynamics is considered, and one model can be related to the other by expressing the injection parameter $s$ in terms of $q$. The relation between $s$ and $q$ in various circumburst medium can be described as \citep{gao13}
\begin{equation}
  s=\frac{10-3k-7q+2kq}{2+q-k}, \quad q=\frac{10-2s-3k+ks}{7+s-2k}.
\end{equation}
For the wind case and the ISM case, the relations between $s$ and $q$ are $s=\frac{4-3q}{q}$ ($q=\frac{4}{3+s}$) and $s=\frac{10-7q}{2+q}$ ($q=\frac{10-2s}{7+s}$), respectively \citep{zhang06}. The Closure relations with $s$ are listed in Table 2. One can obtained the relation between $\alpha$ and $\beta$ in the range $\nu_m < \nu_O < \nu_X < \nu_c$, with $\alpha_{\rm wind}=\frac{2(1+\beta)}{3+s}+\beta$ and $\alpha_{\rm ISM}=\frac{(5-s)(2+\beta)}{7+s}+(\beta-1)$. For GRB 140423A, $\alpha_{\rm wind}=\alpha_{\rm O,II}\ (\rm{or}\ \alpha_{X,II})\approx1.78$, $\alpha_{\rm ISM}=\alpha_{\rm O,III}\ (\rm{or}\ \alpha_{X,III})\approx1.13$, and $p=2\beta+1=2.9$, we obtain an energy injection parameter $s=1.5$, which is located in a reasonable range predicted by the model above.

Based on the above results, there exists a circumburst medium transtion from wind to ISM in GRB~140423A; the transition time $T_{t} \approx 5000$ s. We further present quantitative estimates. The slow cooling with energy injection and $p>2$ in the spectral regime $\nu_m < \nu < \nu_c$ \cite[][]{gao13} gives the expressions of $\nu_m$, $\nu_c$, and $F_{\nu,\max}$ with the $q$ parameter. Here we replace $q$ with $s$. For the wind case,
\begin{equation}
\nu_m=7.0 \times 10^{17}\ \mathrm{Hz}\ (\frac{1+z}{2})^{2/(3+s)} E_{K,52}^{1/2} \epsilon_{e,-1}^{2} \epsilon_{B,-2}^{1/2} t^{-1-2/(3+s)},
\end{equation}
\begin{equation}
\nu_c=5.8 \times 10^{15}\ \mathrm{Hz}\ (\frac{1+z}{2})^{2/(3+s)-2} E_{K,52}^{1/2} A_{*,-1}^{-2} \epsilon_{B,-2}^{-3/2} t^{1-2/(3+s)},
\end{equation}
\begin{equation}
F_{\nu,\max}=4.9 \times 10^{5}\ \mathrm{\mu Jy}\ (\frac{1+z}{2})^{4/(3+s)+2} E_{K,52}^{1/2} A_{*,-1} \epsilon_{B,-2}^{1/2} D_{L,28}^{-2} t^{-2/(3+s)},
\end{equation}
where $E_{K}$ is the kinetic energy of the afterglow, $t$ is the time in seconds since trigger in the observer's frame, $\epsilon_{\rm e}$ is the ratio of shock energy to electron energy, $\epsilon_{\rm B}$ is the ratio of shock energy to magnetic field energy, and $A_\ast$ is the stellar wind density. The minimum frequency decreases with time, $\nu_m \propto t^{-1-2/(3+s)}$, while the cooling frequency increases with time, $\nu_c \propto t^{1-2/(3+s)}$. We have the boundary conditions $\nu_{m}(t_{\rm peak}-T_0)<\nu_{O}$ and $\nu_{c}(t_{\rm peak}-T_0)>\nu_{X}$, where the optical afterglow initial decreases with $\alpha_{\rm O,II} \approx 1.78$ and  $t_{\rm peak}-T_0=206$ s. For the slow cooling, one also can obtain $F_{\nu,O}(t_{\rm peak}-T_0)<F_{\nu,\max}(t_{\rm peak}-T_0)$.
The typical frequency of the optical band is $\sim 10^{14}$~Hz, and the maximum detectable energy of the \emph{Swift}/XRT is 10~keV, or $\sim 10^{18}$~Hz. Thus, we have
\begin{equation}\label{ineq1}
E_{K,52}^{1/2} \epsilon_{e,-1}^{2} \epsilon_{B,-2}^{1/2}<0.22,
\end{equation}
\begin{equation}\label{ineq2}
E_{K,52}^{1/2} A_{*,-1}^{-2} \epsilon_{B,-2}^{-3/2}>28.97,
\end{equation}
\begin{equation}\label{ineq3}
E_{K,52}^{1/2} A_{*,-1} \epsilon_{B,-2}^{1/2}>0.02.
\end{equation}

For the ISM case with energy injection, one has
\begin{equation}
\nu_m=1.37 \times 10^{18}\ \mathrm{Hz}\ \hat{z}^{12/(7+s)-1} E_{K,52}^{1/2} \epsilon_{e,-1}^2 \epsilon_{B,-2}^{1/2} t^{-12/(7+s)},
\end{equation}
\begin{equation}
\nu_c=9.2 \times 10^{18}\ \mathrm{Hz}\ \hat{z}^{1-12/(7+s)} E_{K,52}^{-1/2} n_0^{-1} \epsilon_{B,-2}^{-3/2} t^{12/(7+s)-2},
\end{equation}
\begin{equation}
F_{\nu,\max}=1.1 \times 10^4\ \mathrm{\mu Jy}\ \hat{z}^{24/(7+s)-2} E_{K,52} n_0^{1/2} \epsilon_{B,-2}^{1/2} D_{L,28}^{-2} t^{3-24/(7+s)},
\end{equation}
where $n_{0}$ is the ISM density. In the ISM case, the minimum frequency and the cooling frequency decrease with time, $\nu_m \propto t^{-12/(7+s)}$ and $\nu_c \propto t^{12/(7+s)-2}$ (respectively). Thus, we have the boundary conditions $\nu_{m}(T_{t})<\nu_{O}$ and $F_{\nu,O}(T_{t})<F_{\nu,\max}(T_{t})$ around 5000~s when the optical afterglow transitions from the wind to the ISM, and $\nu_{c}(t)>\nu_{X}$ at the last observational data point of the X-ray afterglow with $t\approx 5\times10^{5}$~s. We then have
\begin{equation}\label{ineq4}
E_{K,52}^{1/2} \epsilon_{e,-1}^2 \epsilon_{B,-2}^{1/2}<8.79,
\end{equation}
\begin{equation}\label{ineq5}
  E_{K,52}^{-1/2} n_0^{-1} \epsilon_{B,-2}^{-3/2}>341.38,
\end{equation}
\begin{equation}
E_{K,52} n_0^{1/2} \epsilon_{B,-2}^{1/2}>0.08.
\end{equation}\label{ineq6}

The wind-to-ISM transition time $T_r$ at transition radius $R_t$ can be estimated as \citep[e.g.,][]{chevalier04,jin09}
\begin{equation}
T_r=1.5 \mathrm{h}\left(\frac{1+z}{2}\right) E_{\mathrm{k}, 53}^{-1} A_{*,-1}^{2} n_{0}^{-1} \approx 5000\rm\ s.
\end{equation}
Thus, we have
\begin{equation}\label{ineq7}
E_{\mathrm{k}, 53}^{-1} A_{*,-1}^{2} n_{0}^{-1} \approx 0.43.
\end{equation}
Based on these limits from the quantitative estimates, we further perform the numerical fit of the multiband afterglow of GRB 140423A.

\section{Modeling: Wind to ISM Transition in the Afterglow}
\label{sec:results}

The external forward shock model with a circumburst medium transition from a stellar wind to the ISM together with energy injection can account for the properties of the GRB 140423A light curve and the spectrum. We adopt the standard external shock model proposed by Sari et al. (1998) and Huang et al. (1999). The spectra in both the $\gamma$-ray and afterglow regions are denoted as $N_e \propto \gamma_e^{-p}$.
The spectral regimes are assumed to be located in $\nu_{\rm m}<\nu<\nu_{\rm c}$, we fix $p=2\beta+1=2.9$. The derived $p$ value of GRB 140423 is larger than the typical one for relativistic shocks owing to first-order Fermi acceleration $p \approx 2.3$ \citep[e.g.,][]{achterberg01,ellison02}. However, it is still consistent with previous studies having a wide distribution \citep[e.g.,][]{curran10,wang15}. The energy injection starting time and ending time are fixed as $t_{\rm s}=10^2$~s, and $t_{\rm e}=10^6$~s. The free parameters of our model are $E_{\rm K,iso}$, $\epsilon_{\rm e}$,  $\epsilon_{\rm B}$, $A_\ast$, $R_t$, $n$, the initial Lorentz factor $\Gamma_{\rm 0}$, the jet opening angle $\theta_j$, and the parameters of the energy injection $L_{\rm 0}$, the index of the density profile $k$, and the index of the energy injection $s$.

The number density of the stellar wind-like medium is given by $n_{\rm wind}(r)=A r^{-k}$, where
\begin{equation}
    A=\frac{\dot M}{4 \pi m_p v_w}=3\times 10^{35} A_\ast \ \rm cm^{-1},
\end{equation}
\noindent
$\dot M$ is the mass-loss rate of the progenitor, $v_w$ is the velocity of the stellar wind, and $A_\ast=(\frac{\dot M}{10^{-5}M_{\odot}})(\frac{1000\rm\ km\ s^{-1}}{v_w})$ \citep[e.g.,][]{Chevalier00,Panaitescu00}.

A Markov Chain Monte Carlo (MCMC) method is adopted to search for the best-fit parameter set. The MCMC is performed using the \texttt{emcee} Python package \citep{emcee}, with 16 walkers running for 10,000 steps in each circle. The preliminary parameters are set in the following ranges: $ E_{\rm K,iso} \in [10^{53}, 10^{56}]$ erg, $\Gamma_0 \in [10, 1000]$, $\log \epsilon_e \in [10^{-3}, 10^{-0.5}]$, $\epsilon_B \in [10^{-8}, 10^{-0.5}]$, $ A_\ast \in [10^{-3}, 10^{3}]$, $R_t \in [10^{16}, 10^{19}]$ cm, $n \in [10^{-3}, 10^{3}]$ cm$^{-3}$, $L_0 \in [10^{51},10^{54}]$ erg s$^{-1}$, $\theta_j \in [0.01,1]$ rad, $k \in [0,3]$, and $s \in [1,4]$.

A set of optimum parameters was obtained in our MCMC results, with $E_{\rm K,iso} = 2.14_{-1.04}^{+1.49}\times10^{55}$ erg, $\Gamma_0 = 162.18_{-75.08}^{+176.66}$, $\epsilon_e = 0.02_{-0.01}^{+0.01}$, $\epsilon_B = 1.66_{-1.37}^{+20.22}\times10^{-6}$, $A_\ast = 1.00_{-0.77}^{+2.31}$, $R_t = 4.07_{-2.08}^{+5.48}\times10^{17}\rm\ cm$, $n = 10.96_{-10.45}^{+133.58} \rm\ cm^{-3}$, $L_0 = 3.09_{-1.54}^{+3.67}\times10^{52} \rm\ erg\ s^{-1}$, $k = 1.98_{-0.02}^{+0.01}$, $s = 1.54_{-0.28}^{+0.25}$. The fitting parameters are consistent with the
statistical properties of a large sample of GRBs \citep[e.g.,][]{wang15}. Since we cannot strictly constrain the jet opening angle through the ``jet break" feature, the lower limit $\theta_j > 0.3$ rad is given. A density jump is supposed to form in the stellar wind to ISM transition \citep{castor75,weaver77}, and here it is $\chi = n/(3\times10^{35}A_\ast R_t^{-2})\approx 6$ appearing at the transition radius $R_t$. Compared with the observational data, Figure 5 shows our best-fit light curves. Figure 6 uses corner plots to show the results of our MCMC parameter estimates.

The GRB radiative efficiency, defined as $\eta_\gamma=E_{\gamma,\rm iso}/(E_{\gamma,\rm iso}+E_{\rm K,iso})$ \citep{Lloyd-Ronning04}, is an essential parameter to probe how efficiently a burst converts its global energy to prompt $\gamma$-ray emission. Our derived value of $\eta_\gamma$ for GRB~140423A is $3.0^{+3.4}_{-1.4} \%$.

\section{Discussion and Conclusions}
\label{sec:dis}

GRB 140423A was detected by {\it Swift}/BAT and {\it Fermi}/GBM. Its broadband afterglow was detected by {\it Swift}/XRT, {\it Swift}/UVOT, and ground-based optical telescopes. We acquired well-sampled optical light curves with the 0.76-m KAIT, the 0.4-m telescope of ISON-NM, the 0.7-m AS-32 of the AAO, the 1-m Zeiss-1000 (East) of the TSHAO, and the 1.6-m AZT-33IK of Mondy from $T_0+$70.96~s to 4.8~d after the {\it Swift}/BAT trigger. This is one more case of prompt optical emission observation. No jet break feature in the optical light curve was detected up to the end of observations on 4.8 days. Our results can be summarized as follows.

 \begin{enumerate}
   \item
Concerning the optical light curves, there exists an onset bump in the early afterglow with a rising index $\alpha_{\rm O,I} = -0.59 \pm 0.04$ peaking at $t_{\rm peak}-T_0 \approx  206 $~s, and then decaying with a steep index $\alpha_{\rm O,II} = 1.78 \pm 0.03$. The optical light curves show a steeper to flatter ``transition" with $\alpha_{\rm O,III} = 1.13 \pm 0.03$ at $T_{t} \sim T_0 + 5000$~s. The observed X-ray afterglow shows an achromatic behavior, as does the optical light curve. There is no apparent evolution of the SED between the X-ray and optical afterglows, with an average value of the photon index $\Gamma \approx 1.95$ ($\beta \approx 0.95$). The temporal and spectral joint fits of the multiwavelength light curves of GRB 140423A are consistent with the external shock model having the circumburst medium transition from a stellar wind to the ISM and including energy injection ($s=1.5$).
   \item
 The GBM time-integrated spectrum can be well fitted with a Band function, with $E_{\rm peak} = 110.0 \pm 16.1$ keV, $\alpha = -0.14 \pm 0.23$, and $\beta = -1.90 \pm 0.08$. The isotropic $\gamma$-ray energy is $E_{\rm \gamma, iso} = (6.54 \pm 0.91) \times 10^{53}$ erg in the 1--10,000 keV band (rest frame), and the GRB radiative efficiency is $\eta_\gamma = 3.0^{3.4}_{1.4} \%$.
   \item
The temporal and spectral joint fits of the multiwavelength light curves of GRB 140423A are consistent with the external shock model having the circumburst medium transition from a stellar wind to the ISM and including energy injection. The best parameters from the MCMC analysis are $E_{\rm K,iso} \approx 2.14\times10^{55}$ erg, $\Gamma_0 \approx 162$, $\epsilon_e \approx 0.02$, $\epsilon_B \approx 1.7\times10^{-6}$, $A_\ast \approx 1.0$, $R_t \approx 4.1\times10^{17}$ cm, $n \approx 11.0 \rm\ cm^{-3}$, $L_0 \approx 3.1\times10^{52} \rm\ erg\ s^{-1}$, $k \approx 1.98$, $s \approx 1.54$. and $\theta_j > 0.3$ rad.

 \end{enumerate}

For the peculiar GRB 140423A, after testing the Closure relation in various afterglow models, we argue that ``transition" from steep to flat around 5000~s was caused by the circumburst medium which transitioned from a stellar wind to the ISM. To satisfy the Closure relation,  energy injection also needs to be included. We derived the parameters of the stellar wind density $A_{\ast} \approx 1.0$, the ISM density $n \approx 11.0 \rm\ cm^{-3}$, and the transition radius of GRB 140423A $R_t \approx 4.1\times10^{17}$ cm. In general, the results for GRB~140423A are consistent with those of previous bursts, such as GRB 050319 \citep{kamble07}, GRB 081109A \citep{jin09}, GRB 160625B \citep{fraija17}, and GRB 190114C \citep{fraija19}. The transition time for GRB 050319 ($T_{t} \approx 1700$~s),  GRB 081109A ($T_{t} \approx 300$~s), and GRB 190114C ($T_{t} \approx 400$~s) happen earlier than for GRB~140423, while that for GRB~160625B is later ($T_{t} \approx 8000$~s). The transition radii are $10^{17}$--$10^{18}$~cm, with $R_t \approx (0.3-1.5) \times 10^{18}$~cm, $4.5\times10^{17}$~cm, $3.1\times10^{18}$~cm, and $2.3\times10^{17}$~cm for GRB 050319, GRB 081109A, GRB 160625B, and GRB 190114C, respectively. The parameters of the circumburst medium vary: GRB 050319, $A_\ast \approx 0.005$--0.032, $n \approx 1$--100~cm$^{-3}$; GRB 081109A, $A_\ast \approx 0.02$, $n \approx 0.12$~cm$^{-3}$; GRB 160625B, $A_\ast \approx 0.2$, $n \approx 10$~cm$^{-3}$; and GRB 190114C, $A_\ast \approx 0.06$, $n \approx 1.1$~cm$^{-3}$.

The $\epsilon_B$ value we find for GRB 140423A is $\sim 1.7 \times10^{-6}$, which is low compared to the values derived from some GRBs \citep[e.g.,][]{PanaitescuKumar2004,McMahon2004,ZhangKobayashi2005,Laskar2013}. However, the $\epsilon_B$ result of GRB 140423A is consistent with recent findings with the $\epsilon_B$ distribution that peaks around $10^{-4}\sim 10^{-5}$ and extends down to $10^{-7}$ \citep[e.g.,][]{kumar09,kumar10,santana14,duran14,wang15,wang18,Beniamini15,Beniamini16,Beniamini17,huangly18}. Our derived radiative efficiency of GRB 140423A is $\sim3\%$, somewhat smaller than typical values in some previous work \citep[e.g.,][]{wang15,Beniamini16}. However, it is consistent with others \citep[e.g.,][]{zhang07} that derived the GRB radiative efficiency based on the low $\epsilon_B$. The low radiative efficiency may be accounted for by the lower values of the magnetic field ($\epsilon_B$); \cite{Beniamini16} show that for $\epsilon_B < 3\times10^{-4}$, the typical prompt radiative efficiency is $\eta_\gamma < 0.2$.

For the circumburst medium of GRB 140423A before the transition time $T_{t}$, our MCMC results show that the index of the density profile $k \approx 1.98$, consistent with the wind-scenario assumption ($k=2$). The prediction of the temporal index for the onset bump rising slope in the circumburst medium is $\alpha=3-[k(\beta+3)/2]$ \citep[e.g.,][]{liang13}. The rising slope of the smooth bump of GRB 140423A ($\alpha_{\rm O,I}=-0.61 \pm 0.04$) is a little shallow compared with the prediction in the wind case with $\alpha=0.03-0.99\beta=-0.91$. This may be caused by that the circumburst medium being more complex in the earlier afterglow \citep[e.g.,][]{dai03,nakar04MNRAS,liang10,liang13,yi13,huangly18,yi20}. Furthermore, the microphysical parameters (e.g., $\epsilon_{e}$ and $\epsilon_{B}$) are suggested to be time-dependent \citep[e.g.,][]{ioka06,huangly18}.

\acknowledgments
We thank the anonymous referee for suggestions that improved this paper, as well as Zi-Gao Dai and Bing Zhang for helpful discussions. This work is supported by the National Natural Science Foundation of China
(grants 11673006, U1938201, 11533003, 11773007, 11851304
and U1938106), the Guangxi Science Foundation (grants
2016GXNSFFA380006, 2017GXNSFBA198206, 2018GXN
SFGA281007, 2017AD22006, 2018GXNSFFA281010, and
2018GXNSFDA281033), the One-Hundred-Talents Program
of Guangxi Colleges, and the High Level Innovation Team and
Outstanding Scholar Program in Guangxi Colleges. A.V.F.'s supernova/GRB group is grateful for financial assistance from the TABASGO Foundation, the Christopher R. Redlich Fund, NASA/{\it Swift} grants NNX12AD73G and 80NSSC19K0156, and the Miller Institute for Basic Research in Science (U.C. Berkeley). KAIT and its ongoing operation were made possible by donations from Sun Microsystems, Inc., the Hewlett-Packard Company, AutoScope Corporation, the Lick Observatory, the US National Science Foundation, the University of California, the Sylvia \& Jim Katzman Foundation, and the TABASGO Foundation. Research at Lick Observatory is partially supported by a generous gift from Google. A.S.P. and  A.A.V. are grateful for RSCF grant 18-12-00378. Observations of the TSHAO observatory are carried out in the framework of Project BR05236322, ``Studies of Physical Processes in Extragalactic and Galactic Objects and Their Subsystems,'' financed by the Ministry of Education and Science of the Republic of Kazakhstan. Observations with the AS-32 telescope of Abastumani Astrophysical Observatory are supported by Shota Rustaveli Science Foundation grant RF-18-1193.  We also acknowledge the use of public data from the \emph{Swift} and {\it Fermi} data archives.


\begin{deluxetable}{cccccc}
\tabletypesize{\small}
\tablewidth{0pt}
\tablecaption{Optical Afterglow Photometry Log of GRB~140423A}
\tablehead{
\colhead{$T-T_{\rm 0}$(mid) (s)\tablenotemark{a}}&
\colhead{Exp. (s)}&
\colhead{Mag\tablenotemark{b}}&
\colhead{$\sigma$\tablenotemark{c}}&
\colhead{Filter}&
\colhead{Telescope}
}
\startdata
\object{	164.96	}&	20	&	13.66	&	0.02	&	$I$	&	KAIT	\\	
\object{	262.96	}&	20	&	13.77	&	0.02	&	$I$	&	KAIT	\\	
\object{	360.96	}&	20	&	13.98	&	0.02	&	$I$	&	KAIT	\\	
\object{	458.96	}&	20	&	14.37	&	0.02	&	$I$	&	KAIT	\\	
\object{	561.96	}&	20	&	14.50	&	0.02	&	$I$	&	KAIT	\\	
\object{	661.96	}&	20	&	14.65	&	0.03	&	$I$	&	KAIT	\\	
\object{	762.96	}&	20	&	14.82	&	0.02	&	$I$	&	KAIT	\\	
\object{	865.96	}&	20	&	15.02	&	0.03	&	$I$	&	KAIT	\\	
\object{	965.96	}&	20	&	15.21	&	0.04	&	$I$	&	KAIT	\\	
\object{	1064.96	}&	20	&	15.38	&	0.04	&	$I$	&	KAIT	\\	
\object{	1167.96	}&	20	&	15.56	&	0.05	&	$I$	&	KAIT	\\	
\object{	1267.96	}&	20	&	15.71	&	0.05	&	$I$	&	KAIT	\\	
\object{	1367.96	}&	20	&	15.84	&	0.05	&	$I$	&	KAIT	\\	
\object{	1468.96	}&	20	&	16.21	&	0.09	&	$I$	&	KAIT	\\	
\object{	1570.96	}&	20	&	16.19	&	0.07	&	$I$	&	KAIT	\\	
\object{	1635.96	}&	20	&	16.28	&	0.07	&	$I$	&	KAIT	\\	
\object{	1700.96	}&	20	&	16.38	&	0.11	&	$I$	&	KAIT	\\	
\object{	1767.96	}&	20	&	16.38	&	0.09	&	$I$	&	KAIT	\\	
\object{	1833.96	}&	20	&	16.37	&	0.08	&	$I$	&	KAIT	\\	
\object{	1903.96	}&	20	&	16.63	&	0.08	&	$I$	&	KAIT	\\	
\object{	1969.96	}&	20	&	16.61	&	0.12	&	$I$	&	KAIT	\\	
\object{	2036.96	}&	20	&	16.70	&	0.09	&	$I$	&	KAIT	\\	
\object{	2273.71	}&	20	&	16.78	&	0.08	&	$I$	&	KAIT	\\	
\object{	2578.36	}&	20	&	16.97	&	0.09	&	$I$	&	KAIT	\\	
\object{	3719.74	}&	20	&	17.66	&	0.20	&	$I$	&	KAIT	\\	
\object{	5027.38	}&	20	&	17.78	&	0.33	&	$I$	&	KAIT	\\	
\object{	18772.99	}&	5880	&	19.50	&	0.30	&	$I$	&	MITSuME	Okayama	\\
\object{	19710	}&	7740	&	19.72	&	0.22	&	$I$	&	MITSuME	Akeno	\\
\object{	31955	}&	7560	&	19.89	&	0.23	&	$I$	&	MITSuME	Akeno	\\
\object{	131.96	}&	20	&	14.85	&	0.02	&	$V$	&	KAIT	\\	
\object{	229.96	}&	20	&	14.79	&	0.02	&	$V$	&	KAIT	\\	
\object{	327.96	}&	20	&	14.90	&	0.02	&	$V$	&	KAIT	\\	
\object{	425.96	}&	20	&	15.24	&	0.03	&	$V$	&	KAIT	\\	
\object{	528.96	}&	20	&	15.51	&	0.03	&	$V$	&	KAIT	\\	
\object{	628.96	}&	20	&	15.55	&	0.03	&	$V$	&	KAIT	\\	
\object{	728.96	}&	20	&	15.76	&	0.04	&	$V$	&	KAIT	\\	
\object{	832.96	}&	20	&	15.89	&	0.04	&	$V$	&	KAIT	\\	
\object{	932.96	}&	20	&	16.21	&	0.05	&	$V$	&	KAIT	\\	
\object{	1030.96	}&	20	&	16.34	&	0.06	&	$V$	&	KAIT	\\	
\object{	1134.96	}&	20	&	16.48	&	0.06	&	$V$	&	KAIT	\\	
\object{	1234.96	}&	20	&	16.84	&	0.08	&	$V$	&	KAIT	\\	
\object{	1334.96	}&	20	&	16.79	&	0.09	&	$V$	&	KAIT	\\	
\object{	1434.96	}&	20	&	16.99	&	0.10	&	$V$	&	KAIT	\\	
\object{	1537.96	}&	20	&	17.13	&	0.10	&	$V$	&	KAIT	\\	
\object{	70.96	}&	1	&	14.78	&	0.03	&	$Clear$	&	KAIT	\\
\object{	74.96	}&	1	&	14.74	&	0.03	&	$Clear$	&	KAIT	\\
\object{	78.96	}&	1	&	14.73	&	0.02	&	$Clear$	&	KAIT	\\
\object{	81.96	}&	1	&	14.64	&	0.02	&	$Clear$	&	KAIT	\\
\object{	84.96	}&	1	&	14.62	&	0.03	&	$Clear$	&	KAIT	\\
\object{	87.96	}&	1	&	14.61	&	0.03	&	$Clear$	&	KAIT	\\
\object{	90.96	}&	1	&	14.52	&	0.03	&	$Clear$	&	KAIT	\\
\object{	94.96	}&	1	&	14.48	&	0.02	&	$Clear$	&	KAIT	\\
\object{	97.96	}&	1	&	14.42	&	0.02	&	$Clear$	&	KAIT	\\
\object{	100.96	}&	1	&	14.41	&	0.02	&	$Clear$	&	KAIT	\\
\object{	196.96	}&	20	&	14.20	&	0.01	&	$Clear$	&	KAIT	\\	
\object{	294.96	}&	20	&	14.29	&	0.01	&	$Clear$	&	KAIT	\\	
\object{	392.96	}&	20	&	14.61	&	0.01	&	$Clear$	&	KAIT	\\	
\object{	495.96	}&	20	&	14.98	&	0.01	&	$Clear$	&	KAIT	\\	
\object{	595.96	}&	20	&	15.04	&	0.01	&	$Clear$	&	KAIT	\\	
\object{	695.96	}&	20	&	15.19	&	0.01	&	$Clear$	&	KAIT	\\	
\object{	799.96	}&	20	&	15.41	&	0.01	&	$Clear$	&	KAIT	\\	
\object{	899.96	}&	20	&	15.59	&	0.02	&	$Clear$	&	KAIT	\\	
\object{	997.96	}&	20	&	15.79	&	0.02	&	$Clear$	&	KAIT	\\	
\object{	1097.96	}&	20	&	16.01	&	0.02	&	$Clear$	&	KAIT	\\	
\object{	1201.96	}&	20	&	16.17	&	0.02	&	$Clear$	&	KAIT	\\	
\object{	1301.96	}&	20	&	16.33	&	0.02	&	$Clear$	&	KAIT	\\	
\object{	1401.96	}&	20	&	16.50	&	0.03	&	$Clear$	&	KAIT	\\	
\object{	1504.96	}&	20	&	16.62	&	0.03	&	$Clear$	&	KAIT	\\	
\object{	1602.96	}&	20	&	16.78	&	0.03	&	$Clear$	&	KAIT	\\	
\object{	1668.96	}&	20	&	16.80	&	0.03	&	$Clear$	&	KAIT	\\	
\object{	1733.96	}&	20	&	16.95	&	0.03	&	$Clear$	&	KAIT	\\	
\object{	1764.98	}&	30	&	17.11	&	0.17	&	$Clear$	&	ISON-NM	\\	
\object{	1800.96	}&	20	&	16.95	&	0.03	&	$Clear$	&	KAIT	\\	
\object{	1829	}&	30	&	17.27	&	0.18	&	$Clear$	&	ISON-NM	\\	
\object{	1869.96	}&	20	&	16.96	&	0.04	&	$Clear$	&	KAIT	\\	
\object{	1916.01	}&	30	&	17.29	&	0.14	&	$Clear$	&	ISON-NM	\\	
\object{	1936.96	}&	20	&	17.04	&	0.03	&	$Clear$	&	KAIT	\\	
\object{	1971.04	}&	30	&	17.18	&	0.17	&	$Clear$	&	ISON-NM	\\	
\object{	2003.96	}&	20	&	17.17	&	0.04	&	$Clear$	&	KAIT	\\	
\object{	2069.96	}&	20	&	17.26	&	0.05	&	$Clear$	&	KAIT	\\	
\object{	2127	}&	60	&	17.59	&	0.11	&	$Clear$	&	ISON-NM	\\	
\object{	2139.96	}&	20	&	17.28	&	0.04	&	$Clear$	&	KAIT	\\	
\object{	2206.96	}&	20	&	17.27	&	0.05	&	$Clear$	&	KAIT	\\	
\object{	2273.96	}&	20	&	17.38	&	0.05	&	$Clear$	&	KAIT	\\	
\object{	2273.96	}&	60	&	17.52	&	0.09	&	$Clear$	&	ISON-NM	\\	
\object{	2339.96	}&	20	&	17.35	&	0.04	&	$Clear$	&	KAIT	\\	
\object{	2403.04	}&	60	&	17.77	&	0.14	&	$Clear$	&	ISON-NM	\\	
\object{	2406.96	}&	20	&	17.50	&	0.04	&	$Clear$	&	KAIT	\\	
\object{	2477.96	}&	20	&	17.56	&	0.05	&	$Clear$	&	KAIT	\\	
\object{	2544.96	}&	20	&	17.61	&	0.06	&	$Clear$	&	KAIT	\\	
\object{	2571.52	}&	90	&	17.76	&	0.12	&	$Clear$	&	ISON-NM	\\	
\object{	2611.96	}&	20	&	17.64	&	0.05	&	$Clear$	&	KAIT	\\	
\object{	2678.96	}&	20	&	17.73	&	0.07	&	$Clear$	&	KAIT	\\	
\object{	2750.96	}&	20	&	17.75	&	0.05	&	$Clear$	&	KAIT	\\	
\object{	2777.96	}&	20	&	17.66	&	0.06	&	$Clear$	&	KAIT	\\	
\object{	2797.46	}&	90	&	18.20	&	0.12	&	$Clear$	&	ISON-NM	\\	
\object{	2976.48	}&	90	&	18.02	&	0.12	&	$Clear$	&	ISON-NM	\\	
\object{	3199.48	}&	120	&	18.09	&	0.16	&	$Clear$	&	ISON-NM	\\	
\object{	3483.96	}&	20	&	18.10	&	0.07	&	$Clear$	&	KAIT	\\	
\object{	3536.52	}&	120	&	18.01	&	0.13	&	$Clear$	&	ISON-NM	\\	
\object{	3550.96	}&	20	&	18.05	&	0.09	&	$Clear$	&	KAIT	\\	
\object{	3617.96	}&	20	&	18.15	&	0.10	&	$Clear$	&	KAIT	\\	
\object{	3684.96	}&	20	&	18.28	&	0.11	&	$Clear$	&	KAIT	\\	
\object{	3754.96	}&	20	&	18.35	&	0.11	&	$Clear$	&	KAIT	\\	
\object{	3773	}&	120	&	18.15	&	0.15	&	$Clear$	&	ISON-NM	\\	
\object{	3821.96	}&	20	&	18.15	&	0.10	&	$Clear$	&	KAIT	\\	
\object{	3888.96	}&	20	&	18.31	&	0.09	&	$Clear$	&	KAIT	\\	
\object{	3954.96	}&	20	&	18.13	&	0.08	&	$Clear$	&	KAIT	\\	
\object{	4024.96	}&	20	&	18.33	&	0.12	&	$Clear$	&	KAIT	\\	
\object{	4027.02	}&	120	&	18.58	&	0.15	&	$Clear$	&	ISON-NM	\\	
\object{	4090.96	}&	20	&	18.41	&	0.08	&	$Clear$	&	KAIT	\\	
\object{	4157.96	}&	20	&	18.44	&	0.12	&	$Clear$	&	KAIT	\\	
\object{	4224.96	}&	20	&	18.42	&	0.12	&	$Clear$	&	KAIT	\\	
\object{	4291.96	}&	20	&	18.27	&	0.13	&	$Clear$	&	KAIT	\\	
\object{	4332.96	}&	150	&	18.30	&	0.17	&	$Clear$	&	ISON-NM	\\	
\object{	4357.96	}&	20	&	18.58	&	0.12	&	$Clear$	&	KAIT	\\	
\object{	4427.96	}&	20	&	18.40	&	0.12	&	$Clear$	&	KAIT	\\	
\object{	4492.96	}&	20	&	18.47	&	0.14	&	$Clear$	&	KAIT	\\	
\object{	4558.96	}&	20	&	18.60	&	0.15	&	$Clear$	&	KAIT	\\	
\object{	4623.96	}&	20	&	18.59	&	0.12	&	$Clear$	&	KAIT	\\	
\object{	4690.96	}&	20	&	18.58	&	0.13	&	$Clear$	&	KAIT	\\	
\object{	4759.96	}&	20	&	18.50	&	0.14	&	$Clear$	&	KAIT	\\	
\object{	4825.96	}&	20	&	18.59	&	0.15	&	$Clear$	&	KAIT	\\	
\object{	5151.51	}&	540	&	18.63	&	0.10	&	$Clear$	&	ISON-NM	\\	
\object{	5161.56	}&	20	&	18.79	&	0.09	&	$Clear$	&	KAIT	\\	
\object{	5466.46	}&	20	&	18.66	&	0.08	&	$Clear$	&	KAIT	\\	
\object{	5649.82	}&	20	&	18.67	&	0.07	&	$Clear$	&	KAIT	\\	
\object{	5847.03	}&	540	&	18.83	&	0.11	&	$Clear$	&	ISON-NM	\\	
\object{	5871.18	}&	20	&	18.65	&	0.06	&	$Clear$	&	KAIT	\\	
\object{	6508.51	}&	540	&	18.83	&	0.16	&	$Clear$	&	ISON-NM	\\	
\object{	7304	}&	720	&	18.90	&	0.11	&	$Clear$	&	ISON-NM	\\	
\object{	8192.02	}&	720	&	19.25	&	0.12	&	$Clear$	&	ISON-NM	\\	
\object{	58872.52	}&	1920	&	$>20.31^{d}$	&	$-$	&	$Clear$	&	AAO	\\
\object{	7920	}&	720	&	19.40	&	0.30	&	$R$	&	Kanata/HONIR	\\	
\object{	19710	}&	7740	&	19.83	&	0.15	&	$R$	&	MITSuME	Akeno	\\
\object{	25145.51	}&	2400	&	20.47	&	0.14	&	$R$	&	Tien Shan	\\	
\object{	31955	}&	7560	&	20.35	&	0.19	&	$R$	&	MITSuME	Akeno	\\
\object{	38945.49	}&	1800	&	20.70	&	0.20	&	$R$	&	Tien Shan	\\	
\object{	62748	}&	300	&	21.29	&	0.06	&	$R$	&	RTT150	\\	
\object{	114292.51	}&	6480	&	22.17	&	0.10	&	$R$	&	Tien Shan	\\	
\object{	204571	}&	5400	&	22.63	&	0.10	&	$R$	&	Mondy	\\	
\object{	288634	}&	5520	&	23.46	&	0.18	&	$R$	&	Mondy	\\	
\object{	369760	}&	6480	&	$>22.90^{d}$	&	$-$	&	$R$	&	Mondy	\\	
\object{	418382	}&	14220	&	$23.90^{e}$	&	0.4	&	$R$	&	Mondy	\\	
\enddata
\tablenotetext{a}{$T-T_{\rm 0}$ is the midpoint of each observation.
The reference time $T_{\rm 0}$ is the \emph{Swift}/BAT burst trigger time.}
\tablenotetext{b}{Not taking into account Galactic extinction.}
\tablenotetext{c}{The uncertainty in the magnitude.}
\tablenotetext{d}{$3\sigma$ upper limit.}
\tablenotetext{e}{Photometry of stacked images obtained on April 27 and 28. $3\sigma$ upper limit is 23.6}

\label{table:obs-data}
\end{deluxetable}

\begin{deluxetable}{cccccccccccc}
\tabletypesize{\tiny}
\tablewidth{0pt}
\tablecaption{Temporal Decay Index $\alpha$ and Spectral Index $\beta$ in Different Spectral Segments.}
\tablehead{
\colhead{}&
\colhead{Spectral regime}&
\colhead{$\beta(p)$}&
\colhead{$\alpha(p)$}&
\colhead{$\alpha(\beta)$}&
\colhead{$\alpha(p,q)$}&
\colhead{$\alpha(p,s)$}&
\colhead{$\alpha(\beta,q)$}&
\colhead{$\alpha(\beta,s)$}&\\
\colhead{}&
\colhead{}&
\colhead{}&
\colhead{(No injection)}&
\colhead{(No injection)}&
\colhead{(Injection)}&
\colhead{(Injection)}&
\colhead{(Injection)}&
\colhead{(Injection)}
}
\startdata
ISM 	&	$\nu_m<\nu<\nu_c$	&	${{p-1 \over 2}}$	&	${3(p-1)\over 4}$	&	${3\beta \over 2}$	&	${(2p-6)+(p+3)q \over 4}$	&	 ${(s+5)p-3(s+1) \over 2(s+3)}$	&	 $(q-1)+\frac{(2+q)\beta}{2}$	&	${s\beta+5\beta-s+1 \over s+3}$	\\
 	&	$\nu>\nu_c$	&	${{p\over 2}}$	&	${3p-2 \over 4}$	&	${3\beta-1 \over 2}$	&	${(2p-4)+(p+2)q\over 4}$	&	${(s+5)p-2(s+1) \over 2(s+3)}$	&	 $\frac{q-2}{2}+\frac{(2+q)\beta}{2}$	&	${s\beta+5\beta-s-1 \over s+3}$	\\
Wind 	&	$\nu_m<\nu<\nu_c$	&	${p-1\over 2}$	&	${3p-1\over 4}$	&	${3\beta+1 \over 2}$	&	${(2p-2)+(p+1)q \over 4}$	&	${6p-s-1 \over s+7}$	&	 $\frac{q}{2}+\frac{(2+q)\beta}{2}$	&	${12\beta-s+5 \over s+7}$	\\
     	&	$\nu>\nu_c$	&	${p\over 2}$	&	${3p-2\over 4}$	&	${3\beta-1 \over 2}$	&	${(2p-4)+(p+2)q\over 4}$	&	${2(3p-s-1) \over s+7}$	&	 $\frac{q-2}{2}+\frac{(2+q)\beta}{2}$	&	${2(6\beta-s-1) \over s+7}$	\\
\enddata
\tablenotetext{a}{$\nu_c$ and $\nu_m$ are the cooling frequency and the characteristic frequency in the synchrotron radiation, respectively.}
\tablenotetext{b}{$p$ is the index of the synchrotron radiating electron spectrum.}
\tablenotetext{c}{$q$ and $s$ is the energy the energy injection parameter.}
\label{table:clousure-relation}
\end{deluxetable}

\clearpage

\begin{figure}
 \centering
\includegraphics[angle=0,scale=0.5]{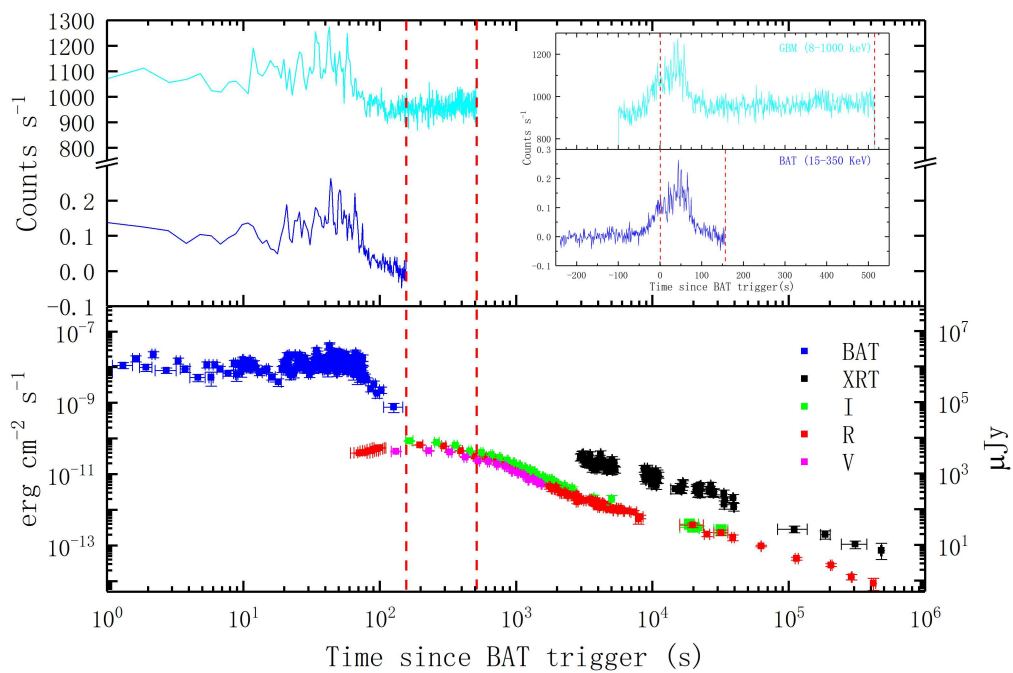}
\caption{Multiband light curves of the prompt emission and afterglow of GRB 140423A on a logarithmic timescale. The inset shows that \emph{Fermi}/GBM and \emph{Swift}/BAT light curves on a linear timescale. The red dashed lines represent the end time of detection of \emph{Fermi}/GBM and \emph{Swift}/BAT, and (in the inset) the zeropoint of time.}
\label{promp-lightcurve}

\end{figure}

\begin{figure}
 \centering
\includegraphics[angle=0,scale=0.5]{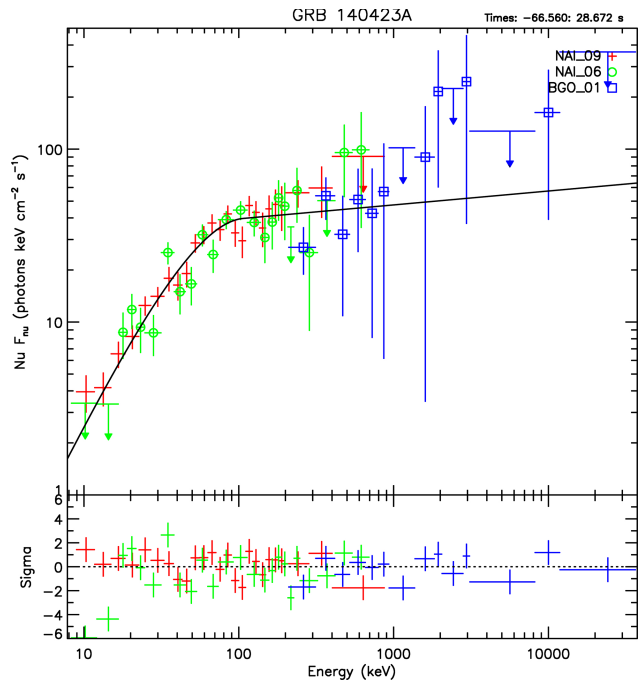}
\caption{The spectral fit for the prompt emission of GRB 140423A with a Band function \citep{band93}.}
\label{promp-spectrum}

\end{figure}

\begin{figure}
 \centering
\includegraphics[angle=0,scale=0.5]{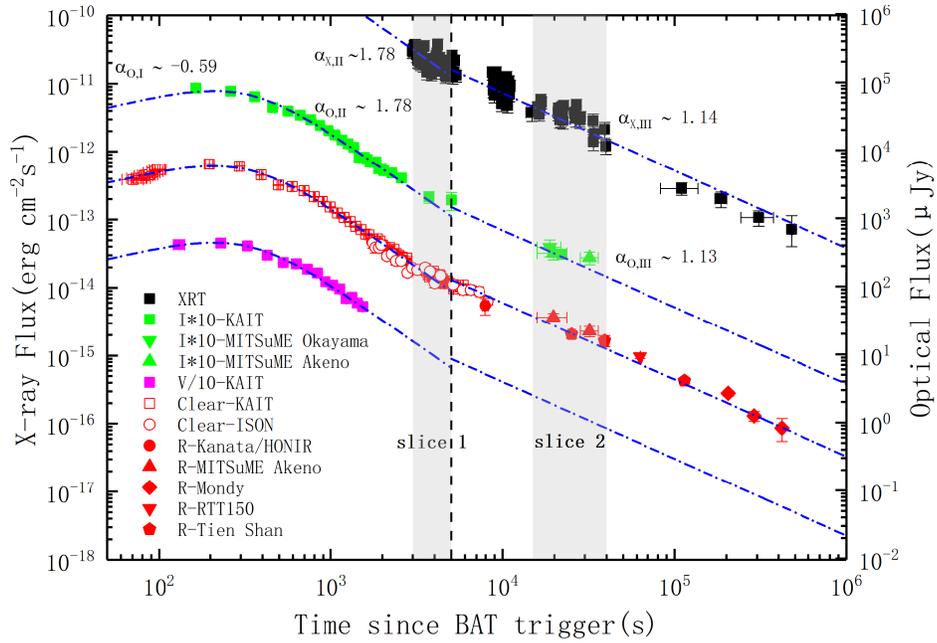}
\caption{The empirical fit to the afterglow light curve. The early-time optical afterglow data obtained with the 0.76-m KAIT at Lick Observatory, and later-time data from other telescopes. The grey zones represent two time slices, 3000--5000~s and 15,000--40,000~s. The vertical black dashed line represents the transition time from stellar wind to ISM.}
\label{Fig-optical-lightcurve}

\end{figure}

\begin{figure}
 \centering
\includegraphics[angle=0,scale=0.5]{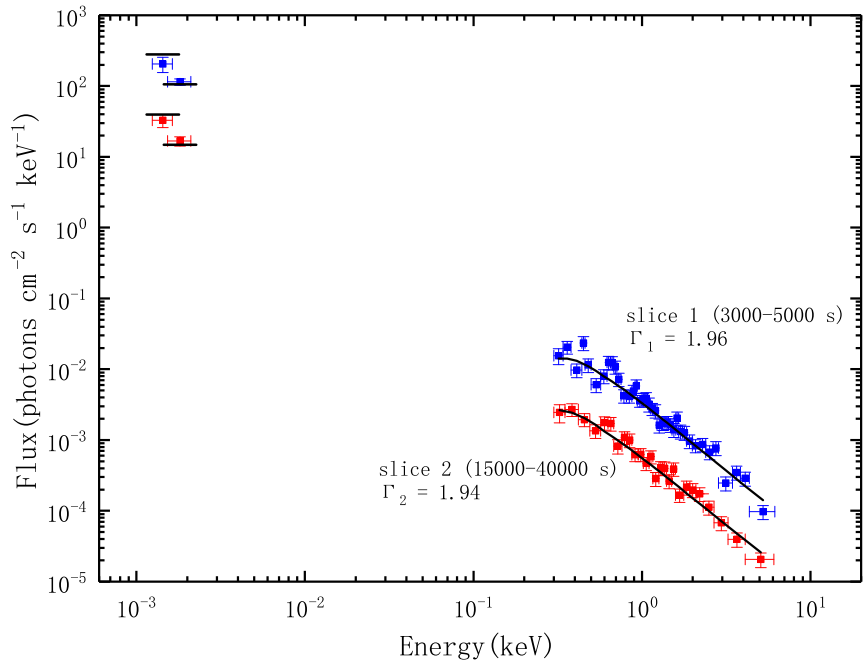}
\caption{Joint spectral fits for the optical and X-ray afterglows, with a single power-law function in two selected time intervals. The fitting results from Xspec are represented with the black solid lines.} \label{Fig-spectral}
\end{figure}

\begin{figure}
 \centering
\includegraphics[angle=0,scale=0.5]{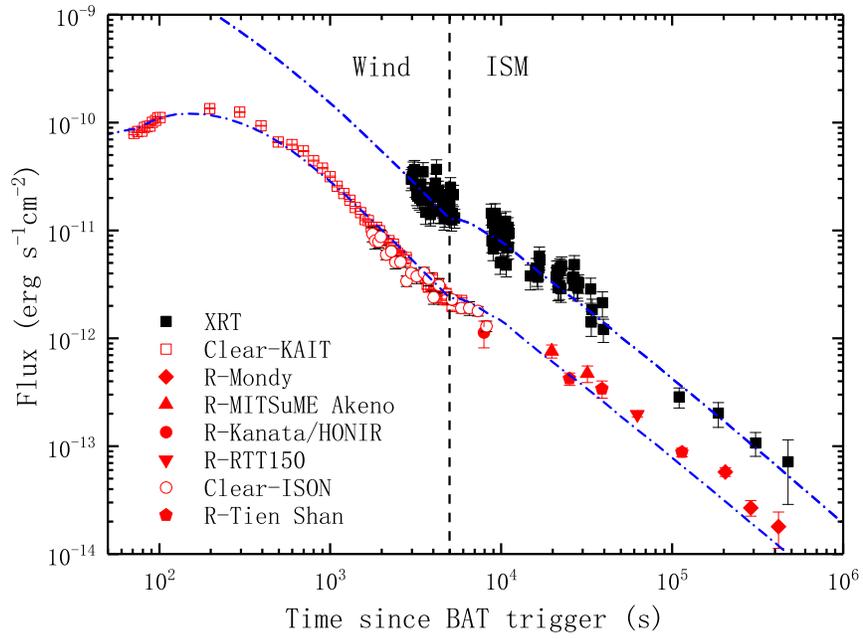}
\caption{Fits of the optical and X-ray afterglow light curves, using the external forward shock model with the circumburst medium transition from stellar wind to ISM and including energy injection.} \label{Fig-modeling}
\end{figure}

\begin{figure}
 \centering
\includegraphics[angle=0,scale=0.28]{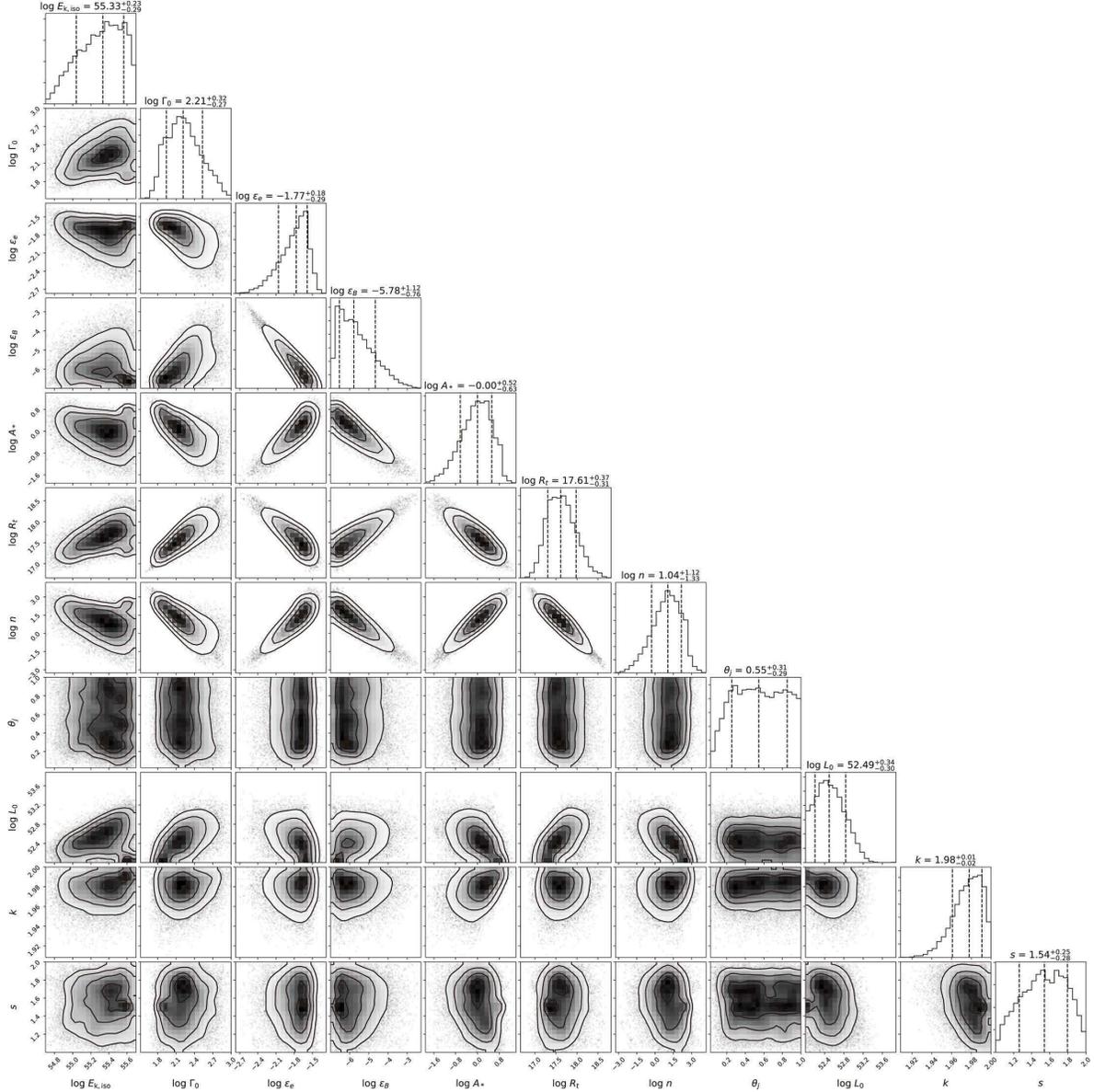}
\caption{A corner plot showing the results of our MCMC parameter estimation for the wind-ISM model, the histograms on the diagonal showing the marginalized posterior densities for each parameter. Two-dimensional projection of the sample is plotted to reveal covariances. The uncertainties are computed as the 16th and 84th percentiles of the posterior samples along each axis, thus representing $1\sigma$ confidence ranges shown with black dashed lines. Our best-fit parameters are also indicated with black dashed lines.} \label{Fig-MCMC}
\end{figure}

\end{document}